\documentclass[preprint,12pt]{elsarticle}

\usepackage{amssymb}
\usepackage{graphicx}
\usepackage{amsmath}
\usepackage{color}

\journal{Speech Communication}

\begin{document}

\begin{frontmatter}



\title{Causal-Anticausal Decomposition of Speech using Complex Cepstrum for Glottal
Source Estimation}


\author[label1]{Thomas Drugman}
\author[label2]{Baris Bozkurt}
\author[label1]{Thierry Dutoit}

 \address[label1]{TCTS Lab, University of Mons, Belgium}
 \address[label2]{Department of Electrical \& Electronics Engineering, Izmir Institute of Technology, Turkey}

\address{}

\begin{abstract}

Complex cepstrum is known in the literature for linearly separating causal and anticausal components. Relying on advances achieved by the Zeros of the Z-Transform (ZZT) technique, we here investigate the possibility of using complex cepstrum for glottal flow estimation on a large-scale database. Via a systematic study of the windowing effects on the deconvolution quality, we show that the complex cepstrum causal-anticausal decomposition can be \emph{effectively} used for glottal flow estimation when specific windowing criteria are met. It is also shown that this complex cepstral decomposition gives similar glottal estimates as obtained with the ZZT method. However, as complex cepstrum uses FFT operations instead of requiring the factoring of high-degree polynomials, the method benefits from a much higher speed. Finally in our tests on a large corpus of real expressive speech, we show that the proposed method has the potential to be used for voice quality analysis.

\end{abstract}


\end{frontmatter}

\section{Introduction}\label{intro}

Glottal source estimation aims at isolating the glottal flow contribution directly from the speech waveform. For this, most of the methods proposed in the literature are based on an inverse filtering process. These methods first estimate a parametric model of the vocal tract, and then obtain the glottal flow by removing the vocal tract contribution via inverse filtering. The methods in this category differ by the way the vocal tract is estimated. In some approaches \cite{ClosedPhase}, \cite{DAP}, this estimation is computed during the glottal closed phase, as the effects of the subglottal cavities are minimized during this period, providing a better way for estimating the vocal tract transfer function. Some other methods (such as \cite{IAIF}) are based on iterative and/or adaptive procedures in order to improve the quality of the glottal flow estimation. Note that a detailed overview of the glottal source estimation methods can be found in various resources such as \cite{Alku-New} or \cite{Walker}.


In this paper we consider a non-parametric decomposition of the speech signal based on the mixed-phase model \cite{Bozkurt-MixedPhase},\cite{Doval-CALM}. According to this model, speech contains a maximum-phase (i.e anticausal) component corresponding to the glottal open phase. In a previous work \cite{Bozkurt-ZZT}, we proposed an algorithm based on the Zeros of the Z-Transform (ZTT) which has the ability to achieve such a deconvolution. However, the ZZT method suffers from high computational load due to the necessity of factorizing large degree polynomials. It has also been discussed in previous studies that the complex cepstrum had the potential to be used for excitation analysis (\cite{Oppenheim},\cite{Quatieri}) but no technique is yet available for reliable glottal flow estimation. This paper more specifically discusses the use of the complex cepstrum for performing the estimation of the glottal open phase from the speech signal, in the light of our previous work on ZZT-based source separation. Almost identical results are obtained with limited computational load, and it is shown that the algorithm is stable enough to enable the analysis of a large database. This manuscript extends our first experiments on such a cepstral decomposition of speech \cite{Drugman-CCD} by providing a more comprehensive theoretical framework, by performing extensive tests on a large real speech corpus and by giving access to a freely available Matlab toolbox.



The goal of this paper is two-fold. First we explain in which conditions complex cepstrum can be used for glottal source estimation. The link with the ZZT-based technique is emphasized and both methods are shown to be two means of achieving the same operation: the causal-anticausal decomposition. However it is shown that the complex cepstrum performs it in a much faster way. Secondly the effects of windowing are studied in a systematic framework. This leads to a set of constraints on the window so that the resulting windowed speech segment exhibits properties described by the mixed-phase model of speech. It should be emphasized that no method is here proposed for estimating the return phase component of the glottal flow signal. As the glottal return phase has a causal character \cite{Doval-CALM}, its contribution is mixed in the also causal vocal tract filter contribution of the speech signal.


The paper is structured as follows. Section \ref{sec:MixedPhaseDecomp} presents the theoretical framework for the causal-anticausal decomposition of voiced speech signals. Two algorithms achieving this deconvolution, namely the Zeros of the Z-Transform (ZZT) and the Complex Cepstrum (CC) based techniques, are described in Section \ref{sec:algos}. The influence of windowing on the causal-anticausal decomposition is investigated in Section \ref{ssec:SynthExp} by a systematic study on synthetic signals. Relying on the conclusions of this study, it is shown in Section \ref{ssec:RealExp} that the complex cepstrum can be efficiently used for glottal source estimation on real speech. Among others we demonstrate the potential of this method for voice quality analysis on an expressive speech corpus. Finally Section \ref{sec:conclu} concludes and summarizes the contributions of the paper.


\section{Causal-Anticausal Decomposition of Voiced Speech}\label{sec:MixedPhaseDecomp}

\subsection{Mixed-Phase Model of Voiced Speech}\label{ssec:MixedPhaseModel}

It is generally accepted that voiced speech results from the excitation of a linear time-invariant system with impulse response $h(n)$, by a periodic pulse train $p(n)$ \cite{Quatieri}:

\begin{equation}\label{eq:Conv}
x(n)=p(n)\star h(n)
\end{equation}

According to the mechanism of voice production, speech is considered as the result of a glottal flow signal filtered by the vocal tract cavities and radiated by the lips. The system transfer function $H(z)$ then consists of the three following contributions: 

\begin{equation}\label{eq:SpeechProd}
H(z)=A\cdot G(z)V(z)R(z)
\end{equation}

where $A$ is the source gain, $G(z)$ the glottal flow over a single cycle, $V(z)$ the vocal tract transmittance and $R(z)$ the radiation load. The resonant vocal tract contribution is generally represented for "pure" vowels by a set of minimum-phase poles ($|v_{2,k}|<1$), while modeling nasalized sounds requires to also consider minimum-phase (i.e causal) zeros ($|v_{1,k}|<1$). $V(z)$ can then be written as the rational form:

\begin{equation}\label{eq:VocalTract}
V(z)=\frac{\prod_{k=1}^{M}(1-v_{1,k} z^{-1})}{\prod_{k=1}^{N}(1-v_{2,k} z^{-1})}
\end{equation}

\begin{figure}[!ht]
  \centering
  \includegraphics[width=0.60\textwidth]{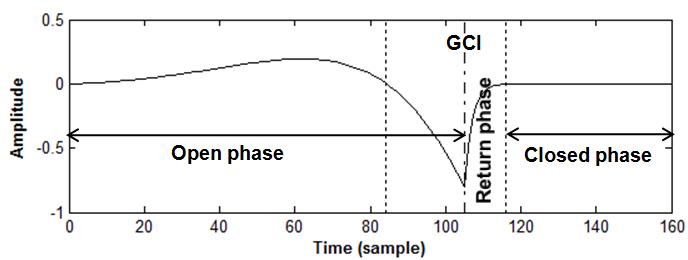}
  \caption{One cycle of a typical waveform of the glottal flow derivative, following the Liljencrants-Fant (LF) model. The different phases of the glottal cycle, as well as the Glottal Closure Instant (GCI) are also indicated.}
  \label{fig:LFmodel}
\end{figure}

During the production of voiced sounds, the airflow evicted by the lungs arises in the trachea and causes a quasi-periodic vibration of the vocal folds \cite{Quatieri}. These latter are then subject to quasi-periodic opening/closure cycles. During the \emph{open phase}, vocal folds are progressively displaced from their initial state because of the increasing subglottal pressure \cite{Childers}. When the elastic displacement limit is reached, they suddenly return to this position during the so-called \emph{return phase}. Figure \ref{fig:LFmodel} displays one cycle of a typical waveform of the glottal flow derivative according to the Liljencrants-Fant (LF) model \cite{Fant}. The limits of these two phases are indicated on the plot, as well as the particular event separating them, called Glottal Closure Instant (GCI).

It has been shown in \cite{Gardner}, \cite{Doval-CALM} that the glottal open phase can be modeled by a pair of maximum-phase (i.e anticausal) poles ($|g_2|>1$) producing the so-called \emph{glottal formant}, while the return phase can be assumed to be a first order causal filter response ($|g_1|<1$) resulting in a \emph{spectral tilt}:

\begin{equation}\label{eq:GlottalFlow}
G(z)=\frac{1}{(1-g_1 z^{-1})(1-g_2 z^{-1})(1-g_2^{\ast} z^{-1})}
\end{equation}

As for the lip radiation, its action is generally assumed as a differential operator:

\begin{equation}\label{eq:LipRadiation}
R(z)=1-rz^{-1}
\end{equation}

with $r$ close to 1. For this reason, it is generally prefered to consider $G(z)R(z)$ in combination, and consequently to study the \emph{glottal flow derivative} or \emph{differentiated glottal flow} instead of the glottal flow itself.

\begin{figure*}[!ht]
  \centering
  \includegraphics[width=1\textwidth]{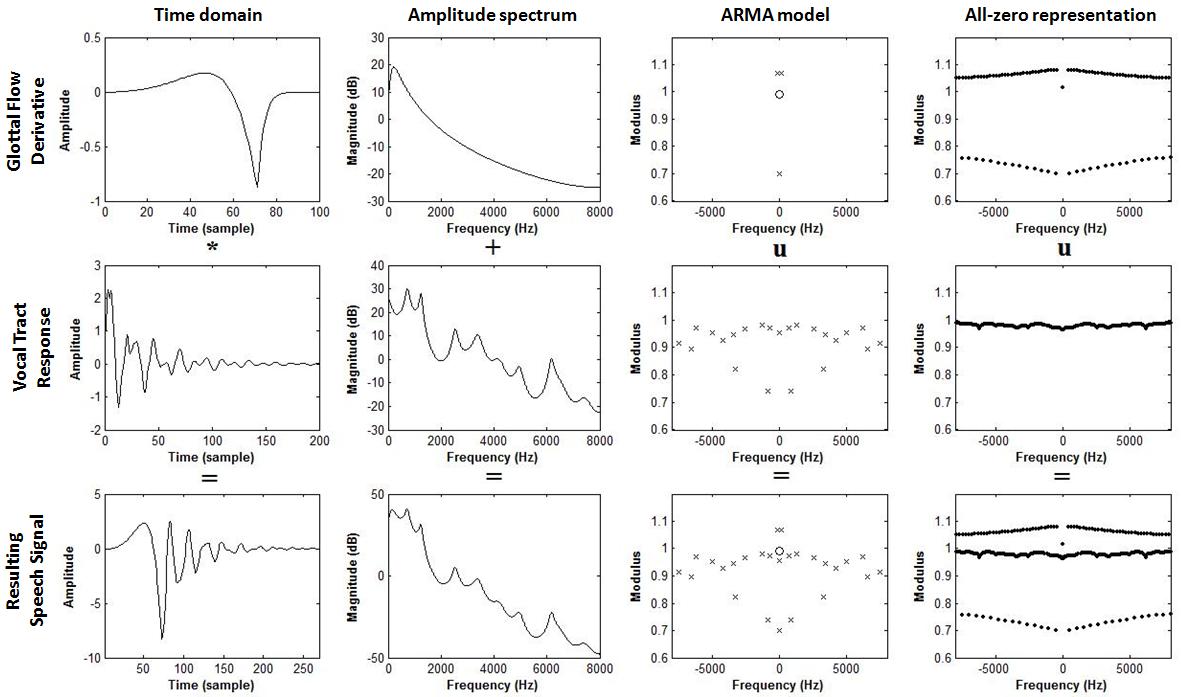}
  \caption{Illustration of the mixed-phase model. The three rows respectively correspond to the glottal flow derivative, the vocal tract response, and the resulting voiced speech. These three signals are all represented in four domains (from the left to the right): waveform, amplitude spectrum, pole-zero modeling, and all-zero (or ZZT) representation. Each column shows how voiced speech is obtained, in each of the four domains.}
  \label{fig:MixedPhaseModel}
\end{figure*}

Gathering the previous equations, the system z-transform $H(z)$ can be expressed as a rational fraction with general form \cite{Oppenheim}:

\begin{equation}\label{eq:TotalH}
H(z)=A\frac{\prod_{k=1}^{M_i}(1-a_k z^{-1})}{\prod_{k=1}^{N_i}(1-b_k z^{-1})\prod_{k=1}^{N_o}(1-c_k z^{-1})}
\end{equation}

where $a_k$ and $b_k$ respectively denote the zeros and poles inside the unit circle ($|a_k|$ and $|b_k|<1$), while $c_k$ are the poles outside the unit circle ($|c_k|>1$). The basic idea behind using causal-anticausal decomposition for glottal flow estimation is the following: \emph{since $c_k$ are only related to the glottal flow, isolating the maximum-phase (i.e anticausal) component of voiced speech should then give an estimation of the glottal open phase}. Besides, if the glottal return phase can be considered as abrupt and if the glottal closure is complete, the anticausal contribution of speech corresponds to the glottal flow. If this is not the case \cite{Stronzo}, these latter components are causal (given their damped nature) and the anticausal contribution of voiced speech still gives an estimation of the glottal open phase.


Figure \ref{fig:MixedPhaseModel} illustrates the mixed-phase model on a single frame of synthetic vowel. In each row the glottal flow and vocal tract contributions, as well as the resulting speech signal, are shown in a different representation space. It should be emphasized here that the all-zero representation (later refered to as the Zeros of Z-Transform (ZZT) representation, and shown in the last column) is obtained by a root finding operation (i.e. a finite($n$)-length signal frame is represented with only zeros in the z-domain). There exists $n-1$ zeros (of the z-transform) for a signal frame with $n$ samples. However the zero in the third row comes from the ARMA model and hence should not be confused with the ZZT. The first row shows a typical glottal flow derivative signal. From the ZZT representation (last column), it can be noticed that some zeros lie outside the unit circle while others are located inside it. The outside zeros correspond to the maximum-phase glottal opening, while the others come from the minimum-phase glottal closure \cite{Bozkurt-ZZT}. The vocal tract response is displayed in the second row. All its zeros are inside the unit circle due to its damped exponential character. Finally the last row is related to the resulting voiced speech. Interestingly its set of zeros is simply the union of the zeros of the two previous components. This is due to the fact that the convolution operation in the time domain corresponds to the multiplication of the z-transform polynomials in the z-domain. For a detailed study of ZZT representation and the mixed-phase speech model, the reader is refered to \cite{Bozkurt-ZZT}.

\subsection{Short-Time Analysis of Voiced Speech}\label{ssec:ShortTime}

For real speech data, Equation (\ref{eq:Conv}) is only valid for a short-time signal \cite{Tribolet}, \cite{Verhelst}. Most practical applications therefore require processing of windowed (i.e short-time) speech segments: 

\begin{align}
s(n)&=w(n)x(n)\label{eq:ConvWin}\\
&=w(n)(A\cdot p(n)\star g(n) \star v(n) \star r(n))
\end{align}

and the goal of the decomposition is to extract the glottal source component $g(n)$ from $s(n)$. As it will be discussed throughout this article, windowing is of crucial importance in order to achieve a correct deconvolution. Indeed, the z-transform of $s(n)$ can be written as:

\begin{align}
S(z)&=W(z)\star X(z)\label{eq:tmp1}\\
&=\sum_{n=0}^{N-1} w(n)x(n)z^{-n}\label{eq:tmp2}\\
&=s(0)z^{-N+1}\prod_{k=1}^{M_i} (z-Z_{C,k}) \prod_{k=1}^{M_o} (z-Z_{AC,k}) \label{eq:ZZT}
\end{align}

where $Z_{C}$ and $Z_{AC}$ are respectively a set of $M_i$ causal ($|Z_{C,k}|<1$) and $M_o$ anticausal ($|Z_{AC,k}|>1$) zeros (with $M_o+M_i=N-1$). As it will be underlined in Section \ref{ssec:ZZT}, Equation \ref{eq:ZZT} corresponds to the ZZT representation.

From these latter expressions, two important considerations have now to be taken into account:
\begin{itemize}
\item Since $s(n)$ is finite length, $S(z)$ is a polynomial in $z$ (see Eq. \ref{eq:ZZT}). This means that the poles of $H(z)$ are now embedded under an all-zero form. Indeed let us consider a single real pole $a$. The z-transform of the related impulse response $y(n)$ limited to $N$ poins is \cite{OppenheimWillsky}:

\begin{equation}
Y(z)=\sum_{n=0}^{N-1} a^n z^{-n}=\frac{1-(az^{-1})^N}{1-az^{-1}}
\end{equation}

which is an all-zero form, since the root of the denominator is also a root of the numerator (and the pole is consequently cancelled).

\item It can be seen from Equations (\ref{eq:tmp1}) and (\ref{eq:tmp2}) that the window $w(n)$ may have a dramatic influence on $S(z)$ \cite{Verhelst}, \cite{Quatieri}. As windowing in the time domain results in a convolution of the window spectrum with the speech spectrum, the resulting change in the ZZT is a highly complex issue to study \cite{Bozkurt-Chirp}. Indeed the multiplication by the windowing function (as in Equation \ref{eq:tmp2}) modifies the root distribution of $X(z)$ in a complex way that cannot be studied analytically. For this reason, the impact of the windowing effects on the mixed-phase model is studied in this paper in an empirical way, as it was done in \cite{Verhelst} and \cite{Quatieri} for the convolutional model.

\end{itemize}

\begin{figure*}[!ht]
  \centering
  \includegraphics[width=1.00\textwidth]{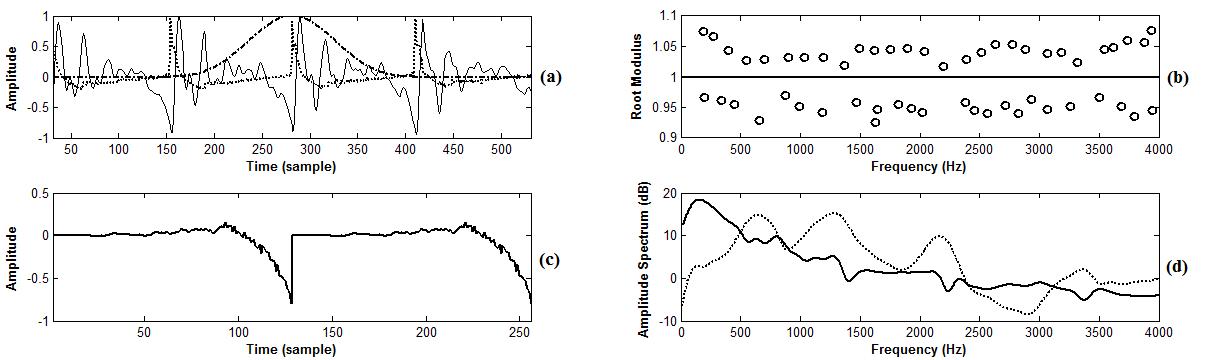}
  \caption{Example of decomposition on a real speech segment using an appropriate window. \emph{(a):} The speech signal (solid line) with the synchronized dEGG (dotted line) and the applied window (dash-dotted line). \emph{(b):} The zero distribution in polar coordinates. \emph{(c):} Two cycles of the maximum-phase component (corresponding to the glottal flow open phase). \emph{(d):} Amplitude spectra of the minimum (dotted line) and maximum-phase (solid line) components of the speech signal. It can be observed that the windowed signal respects the mixed-phase model since the zeros on each side of the unit circle are well separated.}
  \label{fig:GoodDecomp}
\end{figure*}

\begin{figure*}[!ht]
  \centering
  \includegraphics[width=1.00\textwidth]{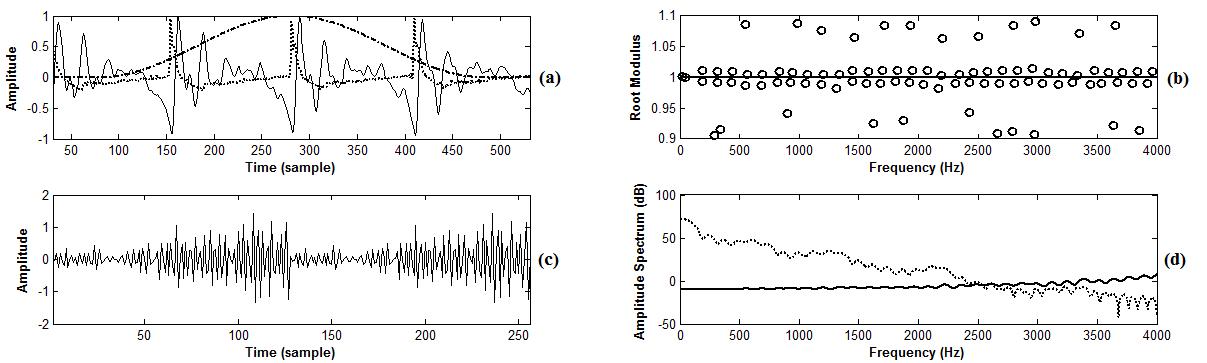}
  \caption{Example of decomposition on a real speech segment using a 25 ms long Hanning window. \emph{(a):} The speech signal (solid line) with the synchronized dEGG (dotted line) and the applied window (dash-dotted line). \emph{(b):} The zero distribution in polar coordinates. \emph{(c):} Two cycles of the maximum-phase component. \emph{(d):} Amplitude spectra of the minimum (dotted line) and maximum-phase (solid line) components  of the speech signal. The zeros on each side of the unit circle are not well separated and the windowed signal does not respect the mixed-phase model. The resulting deconvolved components are irrelevant (while their convolution still gives the input speech signal).}
  \label{fig:BadDecomp}
\end{figure*}

To emphasize the crucial role of windowing, Figures \ref{fig:GoodDecomp} and \ref{fig:BadDecomp} respectively display a case of correct and erroneous glottal flow estimation via causal-anticausal decomposition on a real speech segment. In these figures, the top-left panel (a) contains the speech signal together with the applied window and the synchronized differenced ElectroGlottoGraph $dEGG$ (after compensation of the delay between the laryngograph and the microphone). Peaks in the dEGG signal are informative about the location of the Glottal Closure Instant (GCI). The top-right panel (b) plots the roots of the windowed signal ($Z_{C,k}$ and $Z_{AC,k}$) in polar coordinates. The bottom panels (c) and (d) correspond to the time waveform and amplitude spectrum of the maximum-phase (i.e anticausal) component which is expected to correspond to the glottal flow open phase.

In Figure \ref{fig:GoodDecomp}, an appropriate window respecting the conditions we will derive in Section \ref{ssec:SynthExp} is used. This results in a good separation between the zeros inside and outside the unit circle (see Fig.\ref{fig:GoodDecomp}(b)). The windowed signal then exhibits good mixed-phase properties and the resulting maximum and minimum-phase components corroborate the model exposed in Section \ref{ssec:MixedPhaseModel}. On the contrary, a 25 ms long Hanning window is employed in Figure \ref{fig:BadDecomp}, as widely used in speech processing. It can be seen that even when this window is centered on a GCI, the resulting causal-anticausal decomposition is erroneous. Zeros on each side of the unit circle are not well separated: the windowed signal does not exhibit characteristics of the mixed-phase model. This simple comparison highlights the dramatic influence of windowing on the deconvolution. In Section \ref{ssec:SynthExp}, we discuss in detail the set of properties the window should convey so as to yield a good decomposition.

\section{Algorithms for Causal-Anticausal Decomposition of Voiced Speech}\label{sec:algos}

For a segment $s(n)$ resulting from an appropriate windowing of a voiced speech signal $x(n)$, two algorithms are compared for achieving causal-anticausal decomposition, thereby leading to an estimate $\tilde{g}(n)$ of the real glottal source $g(n)$. The first one relies on the Zeros of the Z-Transform (ZZT, \cite{Bozkurt-ZZT}) and is summarized in Section \ref{ssec:ZZT}. The second technique is based on the Complex Cepstrum (CC) and is described in Section \ref{ssec:CC}. It is important to note that both methods are functionally equivalent to each other, in the sense that they take the same input $s(n)$ and should give the same output $\tilde{g}(n)$. As emphasized in Section \ref{ssec:ShortTime}, the quality of the decomposition then only depends on the applied windowing, i.e whether $s(n)=w(n)x(n)$ exhibits expected mixed-phase properties or not. It will then be shown that both methods lead to similar results (see Section \ref{sssec:SustainedVowel}). However, on a practical point of view, the use of the complex cepstrum is advantageous since it will be shown that it is much faster than ZZT. Note that we made a Matlab toolbox containing these two methods freely available in $http://tcts.fpms.ac.be/$\textasciitilde$drugman/$.



\subsection{Zeros of the Z-Transform-based Decomposition}\label{ssec:ZZT}

According to Equation (\ref{eq:ZZT}), $S(z)$ is a polynomial in $z$ with zeros inside and outside the unit circle. The idea of the ZZT-based decomposition is to isolate the roots $Z_{AC}$ and to reconstruct from them the anticausal component. The algorithm can then be summarized as follows \cite{Bozkurt-ZZT}:

\begin{enumerate}
\item \emph{Window the signal with guidelines provided in Section \ref{ssec:SynthExp},}
\item \emph{Compute the roots of the polynomial $S(z)$,}
\item \emph{Isolate the roots with a modulus greater than 1,}
\item \emph{Compute $\tilde{G}(z)$ from these roots.}
\end{enumerate}

Although very simple, this technique requires the factorization of a polynomial whose order is generally high (depending on the sampling rate and window length). Even though current factoring algorithms are accurate, the time complexity still remains high \cite{FactoringHighDegree}.

In addition to \cite{Bozkurt-ZZT} where the ZZT algorithm is introduced, some recent studies \cite{Sturmel}, \cite{d'Alessandro} have shown that ZZT outperforms other well-known methods of glottal flow estimation in clean recordings. Its main disadvantages are reported as sensitivity to noise and high computational load.

\subsection{Complex Cepstrum-based Decomposition}\label{ssec:CC}

Homomorphic systems have been developed in order to separate non-linearly combined signals \cite{Oppenheim}. As a particular example, the case where inputs are convolved is especially important in speech processing. Separation can then be achieved by a linear homomorphic filtering in the complex cepstrum domain, which interestingly presents the property to map time-domain convolution into addition. In speech analysis, complex cepstrum is usually employed to deconvolve the speech signal into a periodic pulse train and the vocal system impulse response \cite{Quatieri}, \cite{Verhelst}. It finds applications such as pitch detection \cite{PitchDetection}, vocoding \cite{Quatieri-AS}, etc. Based on a previous study \cite{Drugman-CCD}, it is here detailed how to use the complex cepstrum in order to estimate the glottal flow by achieving the causal-anticausal decomposition introduced in Section \ref{ssec:ShortTime}. To our knowledge, no complex cepstrum-based glottal flow estimation method is available in the literature (except this manuscript's introductory version \cite{Drugman-CCD}). Hence it is one of the novel contributions of this paper to introduce one and to test it on a large real speech database.

The complex cepstrum (CC) $\hat{s}(n)$ of a discrete signal $s(n)$ is defined by the following equations \cite{Oppenheim}:

\begin{equation}\label{eq:DFT}
S(\omega)=\sum_{n=-\infty}^{\infty} s(n)e^{-j\omega n}
\end{equation}

\begin{equation}\label{eq:ComplexLog}
\log[S(\omega)]=\log(|S(\omega)|)+j\angle{S(\omega)}
\end{equation}

\begin{equation}\label{eq:ComplexCepstrum}
\hat{s}(n)=\frac{1}{2\pi}\int_{-\pi}^{\pi}{\log[S(\omega)]e^{j\omega n}}\emph{d}\omega
\end{equation}

where Equations (\ref{eq:DFT}), (\ref{eq:ComplexLog}) and (\ref{eq:ComplexCepstrum})  are respectively the Discrete-Time Fourier Transform (DTFT), the complex logarithm and the inverse DTFT (IDTFT). One difficulty when computing the CC lies in the estimation of $\angle{S(\omega)}$, which requires an efficient phase unwrapping algorithm. In this work, we computed the FFT on a sufficiently large number of points (typically 4096) such that the grid on the unit circle is sufficiently fine to facilitate in this way the phase evaluation.

If $S(z)$ is written as in Equation (\ref{eq:ZZT}), it can be easily shown \cite{Oppenheim} that the corresponding complex cepstrum can be expressed as:

\begin{equation}\label{eq:DevComplCepstrum}
\hat{s}(n)= \left\{
\begin{array}{ll}
|s(0)| & \emph{for }n=0\\
\sum_{k=1}^{M_o}{\frac{{Z_{AC,k}}^n}{n}} & \emph{for } n < 0\\
\sum_{k=1}^{M_i}{\frac{{Z_{C,k}}^n}{n}} & \emph{for } n > 0
\end{array} \right.
\end{equation}

This equation shows the close link between the ZZT and the CC-based techniques. Relying on this equation, Steiglitz and Dickinson demonstrated the possibility of computing the complex cepstrum and unwrapped phase by factoring the z-transform \cite{Steiglitz}, \cite{Steiglitz2}. The approach we propose is just the inverse thought process in the sense that our goal is precisely to use the complex cepstrum in order to avoid any factorization. In this way we show that the complex cepstrum can be used as an efficient means to estimate the glottal flow, while circumventing the requirement of factoring polynomials (as it is the case for the ZZT). Indeed it will be shown in Section \ref{sssec:WinFunc} that optimal windows have their length proportional to the pitch period. The ZZT-based technique then requires to compute the roots of generally high-order polynomials (depending on the sampling rate and on the pitch). Although current polynomial factoring algorithms are accurate, the computational load still remains high, with a complexity order of $O(n^2)$ for the fastest algorithms \cite{FactoringHighDegree}, where $n$ denotes the number of samples in the considered frame. On the other hand, the CC-based method just relies on FFT and IFFT operations which can be fast computed, and whose order is $O(N_{FFT}\log(N_{FFT}))$, where $N_{FFT}$ is fixed to 4096 in this work for facilitating phase unwrapping, as mentioned above. For this reason a change in the frame length has little influence on the computation time for the CC-based method. Table \ref{tab:F0OnTime} compares both methods in terms of computation time. The use of the complex cepstrum now offers the possibility of integrating a causal-anticausal decomposition module into a real-time application, which was previously almost impossible with the ZZT-based technique.

\begin{table}[!ht]
\centering
\begin{tabular}{c | c | c}
  & ZZT-based & CC-based \\
Pitch & decomposition & decompostion \\
\hline
60 Hz & 111.4 & 1.038 \\
\hline
180 Hz & 11.2 & 1 \\
\hline
\end{tabular}
\caption{Comparison of the relative computation time (for our Matlab implementation with $F_s=16kHz$) required for decomposing a two pitch period long speech frame. Durations were normalized according to the time needed by the complex cepstrum-based deconvolution for $F_0=180Hz$.}
\label{tab:F0OnTime}
\end{table}

Regarding Equation (\ref{eq:DevComplCepstrum}), it is obvious that causal-anticausal decomposition can be performed using the complex cepstrum, as follows \cite{Drugman-CCD}:
\begin{enumerate}
\item \emph{Window the signal with guidelines provided in Section \ref{ssec:SynthExp},}
\item \emph{Compute the complex cepstrum $\hat{s}(n)$ using Equations (\ref{eq:DFT}), (\ref{eq:ComplexLog}) and (\ref{eq:ComplexCepstrum}),}
\item \emph{Set $\hat{s}(n)$ to zero for $n>0$,}
\item \emph{Compute $\tilde{g}(n)$ by applying the inverse operations of Equations (\ref{eq:DFT}), (\ref{eq:ComplexLog}) and (\ref{eq:ComplexCepstrum}) on the resulting complex cepstrum.}
\end{enumerate}

\begin{figure}[!ht]
  \centering
  \includegraphics[width=0.60\textwidth]{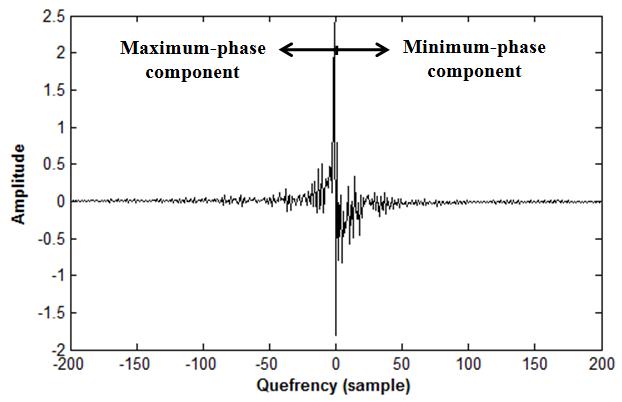}
  \caption{The complex cepstrum $\hat{s}(n)$ of the windowed speech segment $s(n)$ presented in Figure \ref{fig:GoodDecomp}(a). The maximum- (minimum-) phase component can be isolated by only considering the negative (positive) indexes of the complex cepstrum.}
  \label{fig:ComplexCepstrum}
\end{figure}

Figure \ref{fig:ComplexCepstrum} illustrates the complex cepstrum-based decomposition for the example shown in Figure \ref{fig:GoodDecomp}. A simple linear liftering keeping only the negative (positive) indexes of the complex cepstrum allows to isolate the maximum and minimum phase components of voiced speech. It should be emphasized that windowing is very critical as it is the case for the ZZT decomposition. The example in Figure \ref{fig:BadDecomp} (where a $25 ms$ long Hanning window is used) would lead to an unsuccessful decomposition. We think that this critical dependence on the window function, length and location was the main hindrance in developing a complex cepstrum-based glottal flow estimation method, although the potential is known earlier in the literature \cite{Quatieri}. 

It is also worth noting that since the CC method is an alternative means of achieving the mixed-phase decomposition, it suffers from the same noise sensitivity as the ZZT does.

\section{Experiments on Synthetic Speech}\label{ssec:SynthExp}

The goal of this section is to study, on synthetic speech signals, the impact of the windowing effects on the causal-anticausal decomposition. It is one of the main contributions of this study to provide a parametric analysis of the windowing problem and provide guidelines for reliable complex cepstrum-based glottal flow estimation. For this, synthetic speech signals are generated for a wide range of test conditions \cite{Drugman-CCD}. The idea is to cover the diversity of configurations one could find in natural speech by varying all parameters over their whole range. Synthetic speech is produced according to the source-filter model by passing a synthetic train of Liljencrants-Fant (LF) glottal waves \cite{Fant} through an auto-regressive filter extracted by LPC analysis of real sustained vowels uttered by a male speaker. As the mean pitch in these utterances is about 100 Hz, it is reasonable to consider that the fundamental frequency should not exceed 60 and 180 Hz in continuous speech. Experiments in this section can then be seen as a proof of concept on synthetic male speech. Table \ref{tab:Range} summarizes all test conditions.


\begin{table}[!ht]
\centering
\begin{tabular}{c | c}
\hline
Pitch & 60:20:180 Hz \\
\hline
Open quotient & 0.4:0.05:0.9 \\
\hline
Asymmetry coefficient & 0.6:0.05:0.9 \\
\hline
Vowel & /a/, /@/, /i/, /y/ \\
\hline
\end{tabular}
\caption{Table of synthesis parameter variation range.}
\label{tab:Range}
\end{table}

Decomposition quality is assessed through two objective measures \cite{Drugman-CCD}:

\begin{itemize}

\item {\bf Spectral distortion} : Many frequency-domain measures for quantifying the distance between two speech frames have been proposed in the speech coding litterature \cite{SD}. A simple relevant measure between the estimated $\hat{g}(n)$ and the real glottal pulse $g(n)$ is the spectral distortion (SD) defined as \cite{SD}:

\begin{equation}\label{eq:SD1}
SD(g,\hat{g}) = \sqrt{\int_{-\pi}^\pi(20\log_{10}|\frac{G(\omega)}{\hat{G}(\omega)}|)^2\frac{\emph{d}\omega}{2\pi}}
\end{equation}

where $G(\omega)$ and $\hat{G}(\omega)$ denote the DTFT of the original target glottal pulse $g(n)$ and of the estimate $\hat{g}(n)$. To give an idea, it is argued in \cite{Paliwal} that a difference of about 1dB (with a sampling rate of 8kHz) is rather imperceptible.

\item {\bf Glottal formant determination rate} : The amplitude spectrum for a voiced source generally presents a resonance called the \emph{glottal formant} (\cite{Doval-GlottalSpectrum}, see also Section \ref{ssec:MixedPhaseModel}). As this parameter is an essential feature of the glottal open phase, an error
on its determination after decomposition should be penalized. For this, we define the \emph{glottal formant determination rate} as the proportion of frames for which the relative error on the glottal formant frequency is lower than 10\%.

\end{itemize}

This formal experimental protocol allows us to reliably assess our technique and to test its sensivity to various factors influencing the decomposition, such as the window location, function and length. Indeed, Tribolet et al. already observed in 1977 that the window shape and onset may lead to zeros whose topology can be detrimental for accurate pulse estimation \cite{Tribolet}. The goal of this empirical study on synthetic signals is precisely to handle with these zeros close to the unit circle, such that the applied window leads to a correct causal-anticausal separation.

\subsection{Influence of the window location}\label{sssec:WinLocation}

\begin{figure}[!ht]
  \centering
  \includegraphics[width=0.60\textwidth]{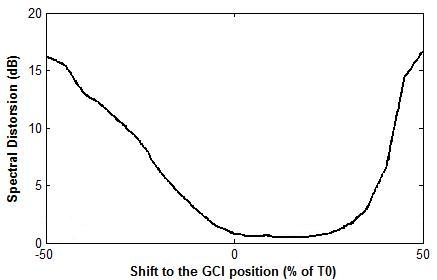}
  \caption{Sensitivity of the causal-anticausal decomposition to a GCI location error. The spectral distortion dramatically increases if a non GCI-centered windowing is applied (particularly on the left of the GCI).}
  \label{fig:ChirpResult}
\end{figure}

In \cite{Quatieri} the need of aligning the window center with the system response is highlighted. Analysis is then performed on windows centered on GCIs, as these particular events demarcate the boundary between the causal and anticausal responses, and the linear phase contribution is removed. Figure \ref{fig:ChirpResult} illustrates the sensitivity of the causal-anticausal decomposition to the window position. It can be noticed that the performance rapidly degrades, especially if the window is centered on the left of the GCI. It is then recommended to apply a GCI-centered windowing. In a concrete application, techniques like the DYPSA algorithm \cite{DYPSA}, or the method we proposed in \cite{Drugman-GCI}, have been shown to give a reliable and accurate estimation of the GCI locations directly from the speech signal. For cases for which GCI information is not available or unreliable, the formalism of the mixed-phase separation has been extended in \cite{chirp} to a chirp analysis, allowing the deconvolution to be achieved in an asynchronous way, but at the expense of a slight performance degradation.


\subsection{Influence of the window function and length}\label{sssec:WinFunc}

\begin{figure}[!ht]
  \centering
  \includegraphics[width=0.60\textwidth]{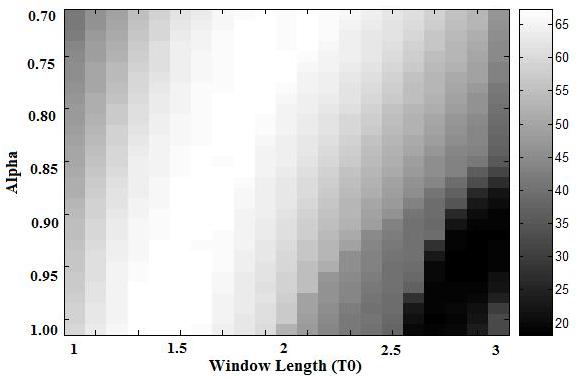}
  \caption{Evolution of the glottal formant determination rate according the window length and shape. Note that the Hanning and Blackman windows respectively correspond to $\alpha=1$ and $\alpha=0.84$.}
  \label{fig:FgDetermRate}
\end{figure}

\begin{figure}[!ht]
  \centering
  \includegraphics[width=0.60\textwidth]{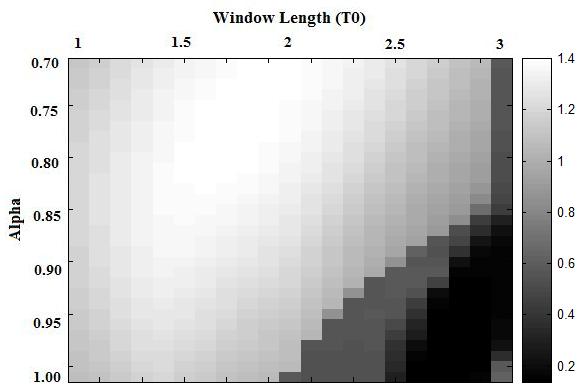}
  \caption{Evolution of the inverse spectral distortion $1/SD$ according the window length and shape. Note that the Hanning and Blackman windows respectively correspond to $\alpha=1$ and $\alpha=0.84$. The inverse $SD$ is plotted instead of the $SD$ itself only for clarity purpose.}
  \label{fig:SD}
\end{figure}

In Section \ref{ssec:ShortTime}, Figures \ref{fig:GoodDecomp} and \ref{fig:BadDecomp} showed an example of correct and erroneous decomposition respectively. The only difference between these figures was the length and shape of the applied windowing. To study this effect let us consider a particular family of windows $w(n)$ of $N$ points satisfying the form \cite{Oppenheim}:

\begin{equation}\label{eq:win}
w(n)=\frac{\alpha}{2}-\frac{1}{2}\cos(\frac{2\pi n}{N-1})+\frac{1-\alpha}{2}\cos(\frac{4\pi n}{N-1})
\end{equation}

where $\alpha$ is a parameter comprised between $0.7$ and $1$ (for $\alpha$ below $0.7$, the window includes negative values which should be avoided). The widely used Hanning and Blackman windows are particular cases of this family for $\alpha=1$ and $\alpha=0.84$ respectively. Figure \ref{fig:FgDetermRate} displays the evolution of the decomposition quality when $\alpha$ and the window length vary. It turns out that a good deconvolution can be achieved as long as the window length is adapted to its shape (or vice versa). For example, the optimal length is about $1.5$ $T_0$ for a Hanning window and  $1.75$ $T_0$ for a Blackman window. A similar observation can be drawn from Figure \ref{fig:SD} according to the spectral distortion criterion. Note that we displayed the inverse spectral distortion $1/SD$ instead of $SD$ only for better viewing purposes. At this point it is interesting to notice that these constraints on the window aiming at respecting the mixed-phase model are sensibly different from those imposed to respect the so-called \emph{convolutional model} \cite{Verhelst}, \cite{Quatieri}. For this latter case, it was indeed recommended to use windows such as Hanning or Hamming with a duration of about 2 to 3 pitch periods. It can be seen from Figure \ref{fig:FgDetermRate} that this would lead to poor causal-anticausal decomposition results. Finally note that it was proposed in \cite{Pedersen} to analytically derive the optimal frame length for the causal-anticausal decomposition, by satisfying an immiscibility criterion based on a Cauchy bound.

\section{Experiments on Real Speech}\label{ssec:RealExp}

The goal of this section is to show that a reliable glottal flow estimation is possible on real speech using the complex cesptrum. The efficiency of this method will be confirmed in Sections \ref{sssec:RealDecomp} and \ref{sssec:SustainedVowel} by analyzing short segments of real speech. Besides we demonstrate in Section \ref{sssec:Emotional} the potential of using complex cepstrum for voice quality analysis on a large expressive speech corpus.

For these experiments, speech signals sampled at $16kHz$ are considered. The pitch contours are extracted using the Snack library \cite{Snack} and the glottal closure instants are located directly from the speech waveforms using the algorithm we proposed in \cite{Drugman-GCI}. Speech frames are then obtained by applying a GCI-centered windowing. The window we use satisfies Equation (\ref{eq:win}) for $\alpha=0.7$ and is two pitch period-long so as to respect the conditions derived in Section \ref{ssec:SynthExp}. Causal-anticausal decomposition is then achieved by the complex cepstrum-based method.

\subsection{Example of Decomposition}\label{sssec:RealDecomp}

\begin{figure*}[!ht]
  \centering
  \includegraphics[width=1.00\textwidth]{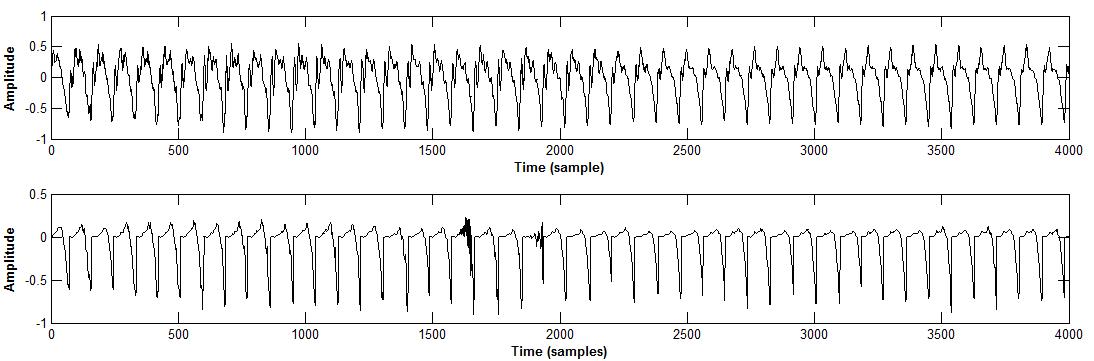}
  \caption{\emph{Top panel:} A segment of voiced speech (diphone \emph{/am/}) uttered by a female speaker. \emph{Bottom panel:} Its corresponding glottal source estimation obtained using the complex cepstrum-based decomposition. It turns out that a reliable estimation can be achieved for most of the speech frames.}
  \label{fig:RealDecomp}
\end{figure*}

Figure \ref{fig:RealDecomp} illustrates a concrete case of decomposition on a voiced speech segment (diphone \emph{/am/}) uttered by a female speaker. It can be seen that even on a nasalized phoneme the glottal source estimation seems to be correctly carried out for most speech frames (i.e the obtained waveforms turn out to corroborate the model of the glottal pulse described in Section \ref{ssec:MixedPhaseModel}). For some rare cases the causal-anticausal decomposition is erroneous and the maximum-phase component contains a high-frequency irrelevant noise. Nevertheless the spectrum of this maximum-phase contribution almost always presents a low-frequency resonance due to the glottal formant. 

\subsection{Analysis of sustained vowels}\label{sssec:SustainedVowel}

\begin{figure*}[!ht]
  \centering
  \includegraphics[width=1.00\textwidth]{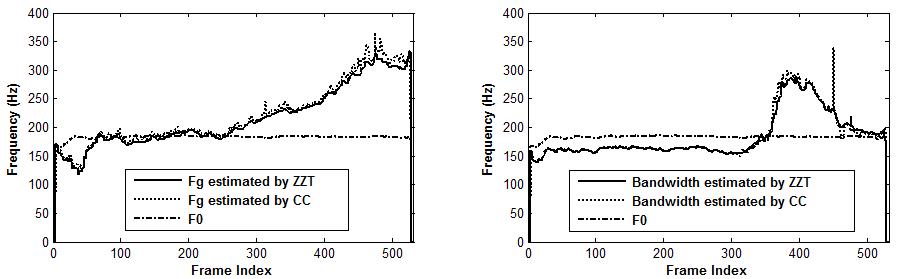}
  \caption{Glottal formant characteristics estimated by both ZZT and CC-based techniques on a real sustained vowel with an increasing pressed effort \cite{Drugman-CCD}. \emph{Left panel:} Evolution of the glottal formant frequency. \emph{Right panel:} Evolution of the glottal formant 3dB bandwidth.}
  \label{fig:RealSpeech}
\end{figure*}

In this experiment, we considered a sustained vowel \emph{/a/} with a flat pitch which was voluntarily produced with an increasing pressed vocal effort. Here the aim is to show that voice quality variation is reflected as expected on the glottal flow estimates obtained using the causal-anticausal decomposition. Figure \ref{fig:RealSpeech} plots the evolution of the glottal formant frequency $Fg$ and bandwidth $Bw$ during the phonation \cite{Drugman-CCD}. These features were estimated with both ZZT and CC-based methods. It can be observed that, as expected, these techniques lead to similar results. The very slight differences may be due to the fact that, for the complex cepstrum, Equation (\ref{eq:DevComplCepstrum}) is realized on a finite number $n$ of points. Another possible explanation is the precision problem in root computation for the ZZT-based technique. In any case, it can be noticed that the increasing vocal effort can be characterized by increasing values of $Fg$ and $Bw$.

\subsection{Analysis of an Expressive Speech Corpus}\label{sssec:Emotional}

The goal of this part is to show that the differences present in the glottal source when a speaker produces various voice qualities can be tracked using causal-anticausal decomposition. For this, the De7 database is used. This database was designed by Marc Schroeder as one of the first attempts of creating diphone databases for expressive speech synthesis \cite{Schroder}. The database contains three voice qualities (modal, soft and loud) uttered by a German female speaker, with about 50 minutes of speech available for each voice quality.

\begin{figure}[!ht]
  \centering
  \includegraphics[width=0.60\textwidth]{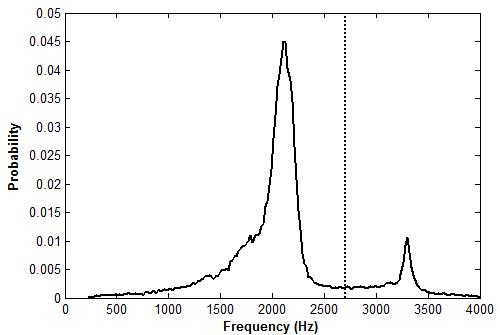}
  \caption{Distribution of the spectral center of gravity of the maximum-phase component, computed for the whole dataset of loud samples. Fixing a threshold around 2.7kHz makes a good separation between correctly and incorrectly decomposed frames.}
  \label{fig:HistoCOG}
\end{figure}

For each voiced speech frame, the complex cepstrum-based decomposition is performed. The resulting maximum-phase component is then downsampled at 8kHz and is assumed to give an estimation of the glottal flow derivative for the considered frame. For each segment of voiced speech, a signal similar to the one illustrated in Figure \ref{fig:RealDecomp} is consequently obtained. For this latter example it was observed that an erroneous decomposition might appear for some frames, leading to an irrelevant high-frequency noise in the estimated anticausal contribution (also observed in Figure \ref{fig:BadDecomp}). One first thing one could wonder is how large is the proportion of such frames over the whole database. As a criterion deciding whether a frame is considered as correctly decomposed or not, we inspect the spectral center of gravity. The distribution of this feature is displayed in Figure \ref{fig:HistoCOG} for the loud voice. A principal mode at around 2kHz clearly emerges and corresponds to the majority of frames for which a correct decomposition is carried out. A second minor mode at higher frequencies is also observed. It is related to the frames where the causal-anticausal decomposition fails, leading to a maximum-phase signal containing an irrelevant high-frequency noise (as explained above). It can be noticed from this histogram (and it was confirmed by a manual verification of numerous frames) that fixing a threshold at around 2750 Hz makes a good distinction between frames that are correctly and incorrectly decomposed. According to this criterion, Table \ref{tab:DecompPerc} summarizes for the whole database the percentage of frames leading to a correct estimation of the glottal flow.

\begin{table}[!ht]
\centering
\begin{tabular}{| c | c |}
\hline
Voice Quality & \% of frames correctly decomposed\\  
\hline
\hline
Loud & $87.22\%$\\
\hline
Modal & $84.41\%$\\
\hline
Soft & $83.69\%$\\
\hline
\end{tabular}
\caption{Proportion of frames leading to a correct causal-anticausal decomposition for the three voice qualities.}
\label{tab:DecompPerc}
\end{table}

For each frame correctly deconvolved, the glottal flow is then characterized by the 3 following common features: 

\begin{itemize}

\item the Normalized Amplitude Quotient($NAQ$): $NAQ$ is a parameter characterizing the glottal closing phase \cite{Alku-NAQ}. It is defined as the ratio between the maximum of the glottal flow and the minimum of its derivative, and then normalized with respect of the fundamental frequency. Its robustness and efficiency to separate different types of phonation was shown in \cite{Alku-NAQ}. Note that a quasi-similar feature called \emph{basic shape parameter} was proposed by Fant in \cite{Fant-Revisited}, where it was qualified as \emph{"most effective single
measure for describing voice qualities"}.

\item the $H1-H2$ ratio: This parameter is defined as the ratio between the amplitudes of the amplitude spectrum of the glottal source at the fundamental frequency and at the second harmonic \cite{Klatt}, \cite{Titze}. It has been widely used as a measure characterizing voice quality \cite{Hansson}, \cite{Fant-Revisited}, \cite{Alku-New}.

\item the Harmonic Richness Factor ($HRF$): This parameter quantifies the amount of harmonics in the magnitude spectrum of the glottal source. It is defined as the ratio between the sum of the amplitudes of harmonics, and the amplitude at the fundamental frequency \cite{Childers}. It was shown to be informative about the phonation type in \cite{Childers2} and \cite{Alku-New}.

\end{itemize}

\begin{figure*}[!ht]
  \centering
  \includegraphics[width=0.95\textwidth]{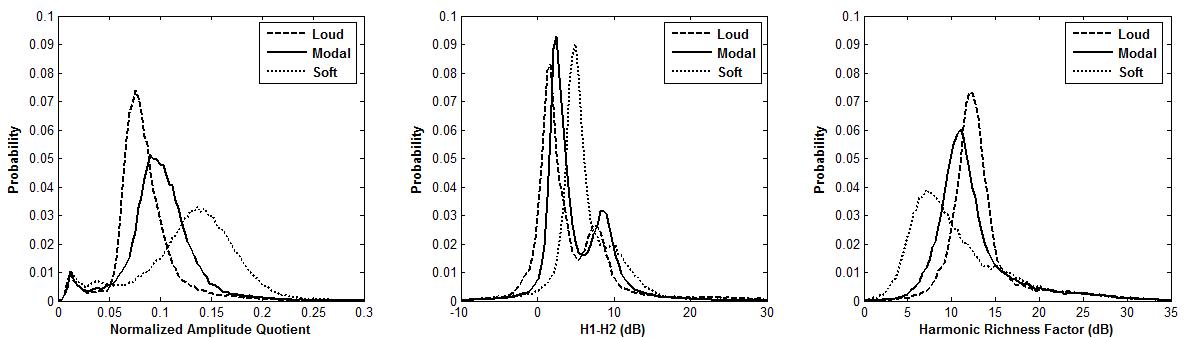}
  \caption{Distributions, computed on a large expressive speech corpus, of glottal source parameters for three voice qualities: \emph{(left)} the Normalized Amplitude Quotient ($NAQ$), \emph{(middle)} the $H1-H2$ ratio, and \emph{(right)} the Harmonic Richness Factor ($HRF$).}
  \label{fig:HistosReal}
\end{figure*}

\begin{figure}[!ht]
  \centering
  \includegraphics[width=0.60\textwidth]{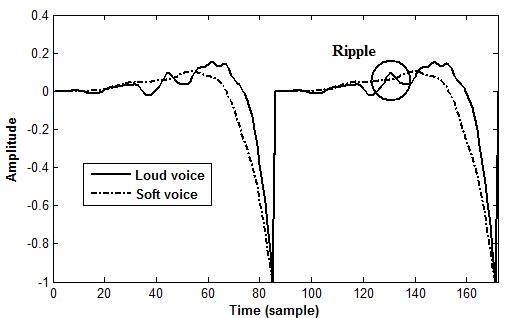}
  \caption{Comparison between two cycles of typical glottal source for both soft (dash-dotted line) and loud voice (solid line). The presence of a ripple in the loud excitation can be observed.}
  \label{fig:Ripple}
\end{figure}

Figure \ref{fig:HistosReal} shows the histograms of these 3 parameters for the three voice qualities. Significant differences between the distributions are observed. Among others it turns out that the production of a louder (softer) voice results in lower (higher) $NAQ$ and $H1-H2$ values, and of a higher (lower) Harmonic Richness Factor ($HRF$). These conclusions corroborate the results recently obtained on sustained vowels by Alku in \cite{Alku-New} and \cite{Alku-NAQ}. Another observation that can be drawn from the histogram of $H1-H2$ is the presence of two modes for the modal and loud voices. This may be explained by the fact that the estimated glottal source sometimes comprises a ripple both in the time and frequency domains \cite{Plumpe}. Indeed consider Figure \ref{fig:Ripple} where two typical cycles of the glottal source are presented for both the soft and loud voice. Two conclusions can be drawn from it. First of all, it is clearly seen that the glottal open phase response for the soft voice is slower than for the loud voice. As it was underlined in the experiment of Section \ref{sssec:SustainedVowel}, this confirms the fact $Fg/F_0$ increases with the vocal effort. Secondly the presence of a ripple in the loud glottal waveform is highlighted. This has two possible origins: an incomplete separation between $Fg$ and the first formant $F_1$ \cite{Bozkurt-Fg}, and/or a non-linear interaction between the vocal tract and the glottis \cite{Plumpe}, \cite{Ananthapadmanabha}. This ripple affects the low-frequency contents of the glottal source spectrum, and may consequently perturb the estimation of the $H1-H2$ feature. This may therefore explain the second mode in the $H1-H2$ histogram for the modal and loud voices (where ripple was observed).

It is also observed that histograms in Figure \ref{fig:HistosReal} present some overlaps. These overlaps may be explained by the three following reasons. \emph{i)} As histograms result from a study led on a large database of connected speech, the glottal production cannot be expected to be perfectly different as a function of the produced voice quality. \emph{ii)} The parametrization of the glottal waveforms by a single feature can only capture a proportion of their differences. \emph{iii)} It might happen for some speech frames that the glottal estimation fails. Although it is impossible to discern and quantify how much each of these causes explains overlaps in Fig. \ref{fig:HistosReal}, we believe that the first two reasons are predominant since irrelevant decompositions have been removed using the spectral criterion.

\section{Discussion and Conclusion}\label{sec:conclu}
This paper explained the causal-anticausal decomposition principles in order to estimate the glottal source directly from the speech waveform. We showed that the complex cepstrum can be effectively used for this purpose as an alternative to the Zeros of the Z-Transform (ZZT) algorithm. Both techniques were shown to be functionally equivalent to each other, while the complex cepstrum is advantageous for its much higher speed, making it suitable for real-time applications. Windowing effects were studied in a systematic way on synthetic signals. It was emphasized that windowing plays a crucial role. More particularly we derived a set of constraints the window should respect so that the windowed signal matches the mixed-phase model. Finally, results on a real speech database (logatoms recorded for the design of an unlimited domain expressive speech synthesizer) were presented for voice quality analysis. The glottal flow was estimated on a large database containing various voice qualities. Interestingly some significant differences between the voice qualities were observed in the excitation. The methods proposed in this paper may be used in several potential applications of speech processing such as emotion detection, speaker recognition, expressive speech synthesis, automatic voice pathology detection and various other applications where real-time glottal source estimation may be useful. Finally note that a Matlab toolbox containing these algorithms is freely available in $http://tcts.fpms.ac.be/$\textasciitilde$drugman/$.

\section*{Acknowledgment}

Thomas Drugman is supported by the Fonds National de la Recherche Scientifique (FNRS). Baris Bozkurt is supported by the Scientific and Technological Research Council of Turkey (TUBITAK). The authors also would like to thank N. Henrich and B. Doval for providing us the speech recording used to create Figure \ref{fig:RealSpeech} and M. Schroeder for the De7 database \cite{Schroder} used in the second experiment on real speech. Authors also would like to thank reviewers for their fruitful feedback.



\bibliographystyle{model1a-num-names}







\end{document}